\documentclass[12pt]{article}
\usepackage{epsfig}

\newcommand\pubnumber{FERMILAB-Conf-00/298-T}
\newcommand\pubdate{\today}
\newcommand\hepnumber{hep-ph/0011181}

\def\csumb{Department of Physics,Carnegie Mellon University, 
Pittsburgh, PA 15213\\
Theory Group, Fermilab, P.O. Box 500, Batavia, IL 60510}
\def\support{\footnote{This work was supported in part by the
U.S. Department of Energy under grant numbers DOE-ER-40682-143 and
DE-AC02-76CH03000.}} 

\def\Title#1{\begin{center} {\Large\bf #1 } \end{center}}
\def\Author#1{\begin{center}{ \sc #1} \end{center}}
\def\Address#1{\begin{center}{ \it #1} \end{center}}

\newcommand\pubblock{\rightline{\begin{tabular}{l} \pubnumber\\
         \pubdate\\ \hepnumber \end{tabular}}}
\newenvironment{Abstract}{\begin{quotation}  }{\end{quotation}}
\newenvironment{Presented}{\begin{quotation} \begin{center} 
             Presented at the\end{center}
      \begin{center}\begin{large}}{\end{large}\end{center} \end{quotation}}
\def\Acknowledgments{\bigskip  \bigskip \begin{center}
          \large\bf Acknowledgments\end{center}}

\makeatletter
\def\section{\@startsection{section}{0}{\z@}{5.5ex plus .5ex minus
 1.5ex}{2.3ex plus .2ex}{\large\bf}}
\def\subsection{\@startsection{subsection}{1}{\z@}{3.5ex plus .5ex minus
 1.5ex}{1.3ex plus .2ex}{\normalsize\bf}}
\def\subsubsection{\@startsection{subsubsection}{2}{\z@}{-3.5ex plus
-1ex minus  -.2ex}{2.3ex plus .2ex}{\normalsize\sl}}

\renewcommand{\@makecaption}[2]{%
   \vskip 10pt
   \setbox\@tempboxa\hbox{\small #1: #2}
   \ifdim \wd\@tempboxa >\hsize     
       \small #1: #2\par          
     \else                        
       \hbox to\hsize{\hfil\box\@tempboxa\hfil}
   \fi}

 \def\citenum#1{{\def\@cite##1##2{##1}\cite{#1}}}
 
\newcount\@tempcntc
\def\@citex[#1]#2{\if@filesw\immediate\write\@auxout{\string\citation{#2}}\fi
  \@tempcnta\z@\@tempcntb\m@ne\def\@citea{}\@cite{\@for\@citeb:=#2\do
    {\@ifundefined
       {b@\@citeb}{\@citeo\@tempcntb\m@ne\@citea\def\@citea{,}{\bf ?}\@warning
       {Citation `\@citeb' on page \thepage \space undefined}}%
    {\setbox\z@\hbox{\global\@tempcntc0\csname b@\@citeb\endcsname\relax}%
     \ifnum\@tempcntc=\z@ \@citeo\@tempcntb\m@ne
       \@citea\def\@citea{,}\hbox{\csname b@\@citeb\endcsname}%
     \else
      \advance\@tempcntb\@ne
      \ifnum\@tempcntb=\@tempcntc
      \else\advance\@tempcntb\m@ne\@citeo
      \@tempcnta\@tempcntc\@tempcntb\@tempcntc\fi\fi}}\@citeo}{#1}}
\def\@citeo{\ifnum\@tempcnta>\@tempcntb\else\@citea\def\@citea{,}%
  \ifnum\@tempcnta=\@tempcntb\the\@tempcnta\else
  {\advance\@tempcnta\@ne\ifnum\@tempcnta=\@tempcntb \else\def\@citea{--}\fi
    \advance\@tempcnta\m@ne\the\@tempcnta\@citea\the\@tempcntb}\fi\fi}
\makeatother

\def\beq{\begin{equation}}
\def\eeq#1{\label{#1}\end{equation}}
\def\eeqn{\end{equation}}

\newenvironment{Eqnarray}%
   {\arraycolsep 0.14em\begin{eqnarray}}{\end{eqnarray}}
\def\beqa{\begin{Eqnarray}}
\def\eeqa#1{\label{#1}\end{Eqnarray}}
\def\eeqan{\end{Eqnarray}}

\let\bar=\overbar

\def\Dslash{\not{\hbox{\kern-4pt $D$}}}
\def\dslash{\not{\hbox{\kern-2pt $\del$}}}

\def\msb{{\bar{\ssstyle M \kern -1pt S}}}

\def\lsim{\mathrel{\raise.3ex\hbox{$<$\kern-.75em\lower1ex\hbox{$\sim$}}}}
\def\gsim{\mathrel{\raise.3ex\hbox{$>$\kern-.75em\lower1ex\hbox{$\sim$}}}}

\begin{document}
\begin{titlepage}
\pubblock

\vfill
\def\thefootnote{\fnsymbol{footnote}}
\Title{$|V_{ub}|$ From Semileptonic Decay and $b\to s\gamma$}
\vfill
\Author{Adam K.\ Leibovich\support}
\Address{\csumb}
\vfill
\begin{Abstract}
Current errors on $|V_{ub}|$ are dominated by model dependence.  For
inclusive decays, the model dependence comes from the Fermi motion of
the $b$ quark.  By combining the endpoint photon and lepton spectra
from the inclusive decays $B\rightarrow X_s\,\gamma$ and $B\rightarrow
X_u\,\ell\,\bar\nu$, it is possible to remove this model dependence.
We show how to combine these rates including the resummation of the
endpoint logs at next to leading order.  The theoretical errors on
$|V_{ub}|$ on the order of 10\% are possible.  We also give a brief
discussion on comparing different extractions.
\end{Abstract}
\vfill
\begin{Presented}
5th International Symposium on Radiative Corrections \\ 
(RADCOR--2000) \\[4pt]
Carmel CA, USA, 11--15 September, 2000
\end{Presented}
\vfill
\end{titlepage}
\def\thefootnote{\arabic{footnote}}
\setcounter{footnote}{0}

\section{Introduction}

The Cabibbo-Kobayashi-Maskawa matrix element $V_{ub}$ is very
important for understanding CP violation in the Standard Model.  An
accurate measurement of $|V_{ub}|$ puts strong constraints on the
Unitarity Triangle.  Unfortunately, $V_{ub}$ is vary hard to measure.
Current measurements have errors that are dominated by model
dependence.  Some of the best extractions so far have come from
exclusive decays, such as $B\to\pi\ell\bar\nu$ or
$B\to\rho\ell\bar\nu$.  The problem with exclusive decays is the
strong hadronic dynamics can not be calculated, and we have to resort
to models, light-cone sum rules, or lattice QCD calculations to obtain
the form factors \cite{ligeti}.  At the present time, all these
methods give around 20\% errors.  A recent measurement from CLEO
\cite{CLEO} using $B\to\rho\ell\bar\nu$ gives $|V_{ub}|=
[3.25\pm0.14{\rm(stat.)}^{+0.21}_{-0.29}{\rm (syst.)}\pm 0.55{\rm
(model)}]\times 10^{-3}$.  In the future, the lattice will give
accurate predictions for the form factors, but until then, a
measurement of 20\% is probably the best we can hope for from
exclusive decays.

In some ways, inclusive decays should provide a straightforward means
to measure $|V_{ub}|$.  All we need to do is measure the total rate
$b\to u\ell\bar\nu$, which is proportional to $|V_{ub}|^2$ and is
known to order $\alpha_s^2$ \cite{vanR}.  If we could measure the
total rate, we would not have to worry about quark-hadron duality
violations, thus a very accurate measurement would be possible.

Unfortunately, there is a very large background from $b\to c$ decays,
which is about 100 times more abundant than $b\to u$ decays.  To
remove this large background, kinematic cuts must be made.  Three
basic cuts are discussed in the literature, each having its own
advantages: a cut on the electron energy spectrum, a cut on the
hadronic invariant mass spectrum, and a cut on the leptonic invariant
mass spectrum.  For now we will concentrate on the electron energy
spectrum, and return to the other cuts later.

\begin{figure}[t]
\begin{center}
\epsfig{file=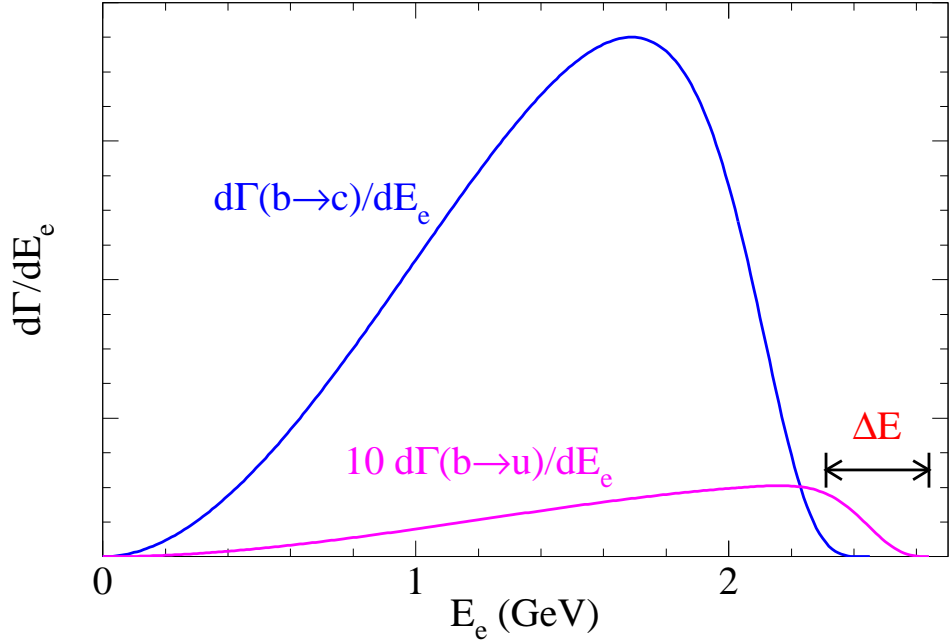,height=4.7in}
\caption[0]{\label{mhtanb} Electron spectrum for semi-leptonic $b$ decay rates to $c$ and
$u$ quarks.  The rate for $b\to ue\bar\nu$ has been multiplied by a
factor of 10.  The region in $\Delta E$ can only have $b\to u$ decays,
and thus is useful for extracting $|V_{ub}|$.}
\label{spectra}
\end{center}
\end{figure}

Since the $u$ quark is much lighter than the $c$ quark, the electron
energy spectrum for $b\to u$ decays extends past the endpoint for
$b\to c$ decays, see Fig.~\ref{spectra}.  Thus it is possible to
remove the charm quark background by cutting above the $b\to c$
endpoint.
All that is necessary is a theoretical prediction for the integrated
rate above the cut.  Unfortunately, putting a cut near the endpoint
introduces a new small mass scale, $\Delta E\sim 300 {\rm\ MeV}$,
which introduces large perturbative $[\log(m_b/\Delta E)]$ and
non-perturbative ($\Lambda/\Delta E)$ corrections.  Therefore both the
perturbative and non-perturbative series must be resummed for the rate
to be trustworthy.

The calculation of the rate begins with the effective Hamiltonian \cite{buras}
\begin{eqnarray}
H_{eff} &=& \frac{-4 G_F}{\sqrt2}V_{ub}(\bar u\gamma_\mu P_L b)
   (\bar\ell\gamma^\mu P_L \nu_\ell) \nonumber\\
&=&\frac{-4 G_F}{\sqrt2}V_{ub}J_\mu J^\mu_\ell,
\end{eqnarray}
obtained by integrating out the $t$ quark and $W$ bosons.  The
differential decay distribution can then be written as the product of
leptonic and hadronic tensors
\begin{equation}
d\Gamma \propto |V_{ub}|^2 L^{\alpha\beta}W_{\alpha\beta}.
\end{equation}
Using the Optical Theorem, the hadronic tensor $W_{\alpha\beta}$ can
be related to the imaginary part of the time ordered product of
currents
\begin{eqnarray}
W_{\alpha\beta} &=& -\frac1\pi {\rm Im}\,T_{\alpha\beta},\\
T_{\alpha\beta} &=& -\frac{i}{2M_B}\int d^4x e^{-iq\cdot x}
  \langle B|T(J_\alpha^\dag(x)J_\beta(0))|B\rangle.
\end{eqnarray}

The time ordered product can be calculated by expanding in an Operator
Product Expansion (OPE).  The Wilson coefficients can be calculated
(over most of phase space) in perturbation theory \cite{CGG,BSUV},
while higher dimensional operators in the OPE are suppressed (over
most of phase space) by powers of $1/m_b$.  Thus, the leading term in
the OPE gives free $b$ quark decay.  The first corrections enter at
order $1/m_b^2$, and are proportional to the Heavy Quark Effective
Theory parameters
\begin{eqnarray}
\lambda_1 &=& \frac{\langle B|\bar h_v(iD)^2h_v|B\rangle}{2 M_B}, \\
\lambda_2 &=& \frac{\langle B|\bar h_vg\sigma_{\mu\nu}
   G^{\mu\nu}|B\rangle}{12 M_B}.
\end{eqnarray}

The problems begin as the energy of the lepton approaches the
endpoint.  Defining $x = 2E_\ell/m_b$ to be the rescaled lepton
energy, the higher dimensional operators in the OPE are actually
suppressed by 
\begin{equation}
\frac\Lambda{m_b(1-x)}\to 1\qquad {\rm as\ } x \to 1.
\end{equation}
Higher dimensional operators in the expansion are no longer
suppressed.  In other words, the expansion is becoming singular as we
approach the endpoint.

The breakdown in the OPE can be seen in the expression for the rate at
order $1/m_b^2$ \cite{NONP},
\begin{equation}
\left.\frac{d\Gamma}{dx}\right|_{O(1/m_b^2)} \propto
   \frac{5\lambda_1+33\lambda_2}{3m_b^2}\theta(1-x)
  -\frac{\lambda_1+33\lambda_2}{6m_b^2}\delta(1-x) 
  -\,\frac{\lambda_1}{6m_b^2}\delta^{\prime}(1-x), 
\end{equation}
by the appearance of singular functions at the endpoint.  

To handle the breakdown of the non-perturbative series, the leading
singular terms must be resummed.  These corrections resum into a
non-perturbative structure function, $f(k_+)$ \cite{N}.  The
differential rate is now a convolution of $f(k_+)$ with the partonic
rate \cite{N,DSU}
\begin{equation}
\label{strucrate}
\frac{d\Gamma}{dE_\ell} = \int dk_+\,f(k_+)
  \frac{d\Gamma_p}{dE_\ell}(m_b^*),
\end{equation}
where $m_b^*=m_b+k_+$.  The structure function is universal, meaning
that the same function occurs for $b\to u\ell\bar\nu$ and $b\to
s\gamma$ decays.  Being a non-perturbative function, $f(k_+)$ is not
known; we do know the first few moments of $f(k_+)$, however.  Thus,
to handle the endpoint region, some model for $f(k_+)$ must be
introduced.  We could in principle extract the structure function from
$b\to s\gamma$ decays and then apply it to $b\to u\ell\bar\nu$, but
this is difficult because of the way $f(k_+)$ enters the rate
(\ref{strucrate}).  Instead, we will skip the step of extracting the
structure function and directly use the $b\to s\gamma$ rate in the
$b\to u\ell\bar\nu$ rate.  But first we need to discuss the
perturbative corrections.

Near the endpoint, the perturbative correction to the rate looks like
\cite{ONEL}
\begin{equation}
\frac{d\Gamma}{dx} \propto  1- \frac{2\alpha_s}{3\pi}
   \left[\log^2(1-x) + \frac{31}{6}\log(1-x) + \pi^2 + \frac54 \right].
\end{equation}
As $x\to1$, the logs become large and the perturbative series breaks
down.  To trust the prediction, the logs need to be resummed.  There
are similarly large logs in the rate for $b\to s\gamma$, so the logs
must be resummed there, too \cite{LR}.

It is possible to resum the series using Infrared Factorization, which
is also used for DIS, Drell-Yan, etc.  The idea is that in the
endpoint region, the light quark is shot out with large energy, but
with small invariant mass.  This quark produces a jet of particles
through collinear radiation.  While the constituents of the jet can
talk to each other (and the original $b$ quark) through soft gluons,
hard gluon exchange is disallowed.  The soft radiation cannot tell if
the jet was initiated by a $u$ quark or an $s$ quark, thus it will be
the same for $b\to u$ and $b\to s$ decays.

Mathematically, there is a separation of momentum regions into \cite{SK}
\begin{eqnarray}
{\rm Hard}\ (H):&& k_+\sim k_-\sim k_t = O(m_b),\\
{\rm Jet}\ (J):&& k_+ = O[m_b(1-x)],\ k_- = O(m_b),
  k_t = O(m_b\sqrt{1-x}),\\
{\rm Soft}\ (S):&& k_+\sim k_-\sim k_t = O[m_b(1-x)].
\end{eqnarray}
By introducing a factorization scale $\mu$ to keep these regions
separated, we can write the rate in factorized form as
\begin{equation}
\frac{d\Gamma}{dx} \sim \int dz S(z,\mu) J(z,\mu) H(\mu).
\end{equation}
The soft function $S(z,\mu)$ is the same for $b\to u$ and $b\to s$,
while $J(z,\mu)$ and $H(\mu)$ depend on the process.

The rate completely factorizes after taking moments, 
\begin{eqnarray}
\label{radmom}
M_N^\gamma &=& \int_o^{M_B/m_b} dx\,x^{n-1}\,
   \frac1{\Gamma_o^\gamma}\frac{d\Gamma^\gamma}{dx} = 
   S_N J_N^\gamma H_N^\gamma,\\
\label{slmom}
M_N^{sl} &=& -\int_o^{M_B/m_b} dx\,x^{n-1}\,
   \frac1{\Gamma_o^\gamma}\frac{d}{dx}\frac{d\Gamma^\gamma}{dx}
   \nonumber\\
&=& \int dx_\nu S_N J_N^{sl}(x_\nu) H_N^{sl}(x_\nu),
\end{eqnarray}
where $M_N^\gamma$ and $M_N^{sl}$ are the moments of the $b\to
s\gamma$ and $b\to u\ell\bar\nu$ rates, respectively.

The soft function contains perturbative and non-perturbative pieces
\begin{equation}
S_N=f_N\sigma_N,
\end{equation}
where $f_N$ are the moments of the structure function introduced earlier.
Thus we can write the moments (\ref{radmom}) and (\ref{slmom}) as
\begin{eqnarray}
\label{radfmom}
M_N^\gamma &=& f_N\sigma_N J_N^\gamma H_N^\gamma,\\
\label{slfmom}
M_N^{sl} &=& \int dx_\nu f_N\sigma_N J_N^{sl}(x_\nu) H_N^{sl}(x_\nu).
\end{eqnarray}

All the large logarithms are contained in the combination
$\sigma_N J_N$.  The only fact we need about the perturbative
resummation is that after resumming, including next-to-leading logarithms,
there is the relation \cite{AR}
\begin{equation}
\label{resumrel}
\sigma_N J_N^{sl} = \sigma_N J_N^{\gamma}\exp[g_{sl}(\alpha_s\log N)],
\end{equation}
where $\exp[g_{sl}(\alpha_s\log N)]$ is a known function.  

We can now combine the above results.  Substituting first
(\ref{resumrel}) into (\ref{slfmom}), and then (\ref{radfmom}) into
the result, we get
\begin{eqnarray}
\label{finalmom}
M_N^{sl} &=& \int dx_\nu f_N \sigma_N J_N^{sl}(x_\nu) H_N^{sl}(x_\nu)
  \nonumber\\
&=& \int dx_\nu f_N \sigma_N J_N^{\gamma} \exp[g_{sl}(\alpha_s\log N)]
   H_N^{sl}(x_\nu)\nonumber\\
&=& \int dx_\nu \frac{M_N^\gamma}{H^\gamma} 
   \exp[g_{sl}(\alpha_s\log N)] H_N^{sl}(x_\nu).
\end{eqnarray}
Note that the dependence on the unknown structure function $f_N$ has
been eliminated.

We can go back to $x$-space by taking an inverse Mellin transform.  The
left-hand side of (\ref{finalmom}) is just the semi-leptonic rate.
The right-hand side is a convolution of the $b\to s\gamma$ rate with a
known function.  Rearranging, we can write this as \cite{LLR1}
\begin{equation}
\frac{|V_{ub}|^2}{|V_{ts}^*V_{tb}|^2} =
 \frac{\int \Gamma(b\to u\ell\nu)}
  {\int \int d\Gamma^\gamma/dx^\gamma
    * K(x^\gamma, \alpha_s)}.
\end{equation}
So in words, what we have done is written $|V_{ub}|^2$ as the ratio of
the $b\to ue\bar\nu$ rate over a convolution of the $b\to s\gamma$
rate with a known function.

What are the uncertainties?  First, there are higher order corrections
that we neglected, which enter at the order of $\Lambda/m_b$,
$\alpha_s(1-x)$ and $(1-x)^3$.  For the value of the electron energy
cut, $x_{\rm cut} \approx 0.87$, these corrections should all be less
that 10\%.  Of course, we are estimating the size of the higher order
corrections, since they have not been calculated.  They may be larger
or smaller by a factor of 2 or 3.  Without calculating the corrections
directly, it is not possible to know.  We will come back to this
qualification shortly.

Second, there are the violations of quark-hadron duality.  These
violations are hard to quantify, but they should be small if we are
not dominated by a just a few resonances; the more final states, the
smaller the duality violations.  In the region that we are interested
in, it does not appear that we are dominated by resonances, so
neglecting them should be okay.  It would be better if we could have a
larger number of the decay products.

This is possible if we cut on different kinematic variables.  The
other variables discussed in the literature are the hadronic invariant
mass \cite{FLW}, and the lepton invariant mass \cite{BLL}.  The
hadronic invariant mass spectrum also has dependence on the structure
function \cite{FLW,MR}, which introduces model dependence.  However,
by using a method analogous to the one described above for the
electron spectrum, the dependence on the structure function can be
eliminated \cite{LLR2,LLR3}.  The errors from higher order corrections
are similar to the electron spectrum and should be around 10\%.  The
main advantage of the hadronic invariant mass is that after a cut to
remove the charm background, between 40\% and 80\% of the possible
final states will be included.  This is much larger than for the
electron spectrum, which includes about 10\% of the possible final
states.  Thus the quark-hadron duality violations should be
negligible.

The leptonic invariant mass cut has different advantages
\cite{BLL,luke}.  Here the structure function is not important, so we
do not need to do anything to remove this model dependence.  Higher
order non-perturbative corrections are on the order of
$(\Lambda/m_c)^3$, which leads to an error again of around 10\%.  This
disadvantage for this cut is the fraction of final states included
after the cut is around 20\%, so the quark-hadron duality errors may
be an issue.  There is also some question about how good a resolution
can be obtained on the lepton invariant mass, which is the only
immediate problem for this method.

All three of the above methods should have a theoretical uncertainty
(modulo quark-hadron duality violations) of around 10\%.  Again, these
are estimates of higher order corrections.  The actual errors may be
bigger or smaller.  Also, the duality violations could enter in
different ways for each measurement.  To really trust any extraction
of $|V_{ub}|$, we should measure it as many ways as possible, and only
after (or if) there is a convergence of the results should we trust
the extracted value.

\Acknowledgments 
I would like to my collaborators I.~Low and I.~Z.~Rothstein, and the
organizers of Radcor 2000 for a very enjoyable conference.

\end{document}